# Electrostatic and Quantum-Transport Engineering of Multilayer Ferroelectric Tunnel Junctions for Improved TER–RA Trade-off

Balram Khattar, Manish Anand, and Abhishek Sharma, *Member, IEEE*

***Abstract*—Ferroelectric tunnel junctions (FTJs) are promising for non-volatile memory and neuromorphic computing, but achieving a high tunnel electroresistance ratio (TER) with a low resistance-area (RA) product remains challenging. In this work, TiN/$Hf_{0.5}Zr_{0.5}O_2$ (HZO)-based FTJs containing zero to three $TiO_2$ and IGZO-based quantum wells (QWs) are analyzed using the Miller ferroelectric (FE) polarization model integrated with the non-equilibrium Green's function (NEGF) formalism to examine the effects of QW material and thickness on polarization and transport. $TiO_2$-based structures exhibit higher polarization and TER than IGZO-based structures. At a read bias of 0.1 V, the uniform $TiO_2$ double-QW structure, with 1-nm-thick QW and insulating layer (IL) layers, gives a TER of $2.71 \times 10^{10}\%$ and an RA product of $5 \times 10^4\ \Omega \cdot$ cm$^2$, compared with $5.95 \times 10^7\%$ and $2 \times 10^5\ \Omega \cdot$ cm$^2$, respectively, for IGZO. Both uniform and non-uniform $TiO_2$ double-QW structures exhibit resonant tunneling (RT) and negative differential resistance (NDR) associated with QW-confined states. Increasing the first QW thickness to 1.38 nm, while keeping the second QW and both IL layers at 1 nm, shifts the confined-state energies and modifies resonant transport. The selected non-uniform structure achieves a TER of $2.95 \times 10^{11}\%$ and an RA product of $1.25 \times 10^3\ \Omega \cdot$ cm$^2$ at a read bias of 0.1 V. Elastic scattering changes the amplitude and width of QW resonant-transmission peaks, with a stronger effect on the low resistance state (LRS) than on the high resistance state (HRS). These results show that QW material and thickness optimization can improve the TER–RA trade-off, while electrode screening and elastic scattering further influence polarization-dependent transport in multilayer FTJs.**

***Index Terms*—tunneling electro-resistance, quantum wells, resonant tunneling, non-volatile memory.**

## I. Introduction

Ferroelectric tunnel junctions (FTJs) offer potential for non-volatile memory (NVM) and neuromorphic computing because their electrical resistance can be controlled through the polarization of a ferroelectric (FE) barrier. Reversing the FE polarization changes the effective tunneling barrier, switching the device between a LRS and a HRS [1], [2]. The difference in current density between these two states is expressed by the TER, defined as

$$\mathrm{TER}(\%) = \left[\frac{J_{LRS}}{J_{HRS}} - 1\right] \times 100, \quad (1)$$

where $J_{LRS}$ and $J_{HRS}$ are the current densities in the LRS and HRS, respectively [3]. $HfO_2$-based FTJs have received considerable attention because of their CMOS compatibility, scalability, and potential for low-energy operation [4]–[6]. However, achieving a high TER together with a low LRS RA product remains challenging, particularly for thicker FE barriers, where increased tunneling resistance limits the read current [7]. Introducing QWs into an asymmetric tunneling barriers provides resonant transport paths when their confined energy levels align with the transport energy window [8], [9]. The resulting transmission in the two polarization states depends on the QW material properties, particularly dielectric permittivity, electron affinity, and effective mass, as well as QW thickness, thereby influencing the TER and RA characteristics.

$TiO_2$ and In-Ga-Zn-O (IGZO) have both previously been integrated into hafnia-based FE structures. $TiO_2$ has been used as a dielectric interlayer in HZO-based FTJs, where its insertion improves FE properties and TER [10]. It has also been used as an interlayer to enhance ferroelectricity in HZO films [11]. IGZO has also been integrated with HZO in ferroelectric devices [12]. In our previous work, we investigated an asymmetric HZO-based FTJ incorporating a single $TiO_2$ QW using a Preisach-based FE polarization model coupled with NEGF method [13]. A single QW FTJ incorporating a $Ta_2O_5$ well was also investigated using the transfer-matrix method (TMM) [14]. Multiple QW hafnia-based FTJs were later analyzed using the Miller polarization model and TMM, assuming ideal electrode screening [15]. Apart from QW material selection, layer thickness also affects the confined-state energies and resonant transport. The IL thickness determines the wave-function penetration and coupling between neighboring QWs, whereas the QW thickness controls the confinement energies and positions of the confined states. Previous studies of superlattice structures have reported changes in transmission with non-uniform layer thicknesses [16], [17]. Such changes can modify resonant transport and lead to NDR [18]. However, a systematic investigation of the influence of QW material properties and thickness configurations on resonant transport and the TER–RA trade-off remains limited in multilayer FTJs. The combined effects

Balram Khattar is with the Department of Electrical Engineering, IIT Ropar, Rupnagar, Punjab 140001, India.
Manish Anand is with the Department of Physics, Bihar National College, Patna University, Patna-800004, India.
Abhishek Sharma is with the School of Computing and Electrical Engineering, Indian Institute of Technology Mandi, Mandi 175005, India, India (e-mail: abhi@iitmandi.ac.in).

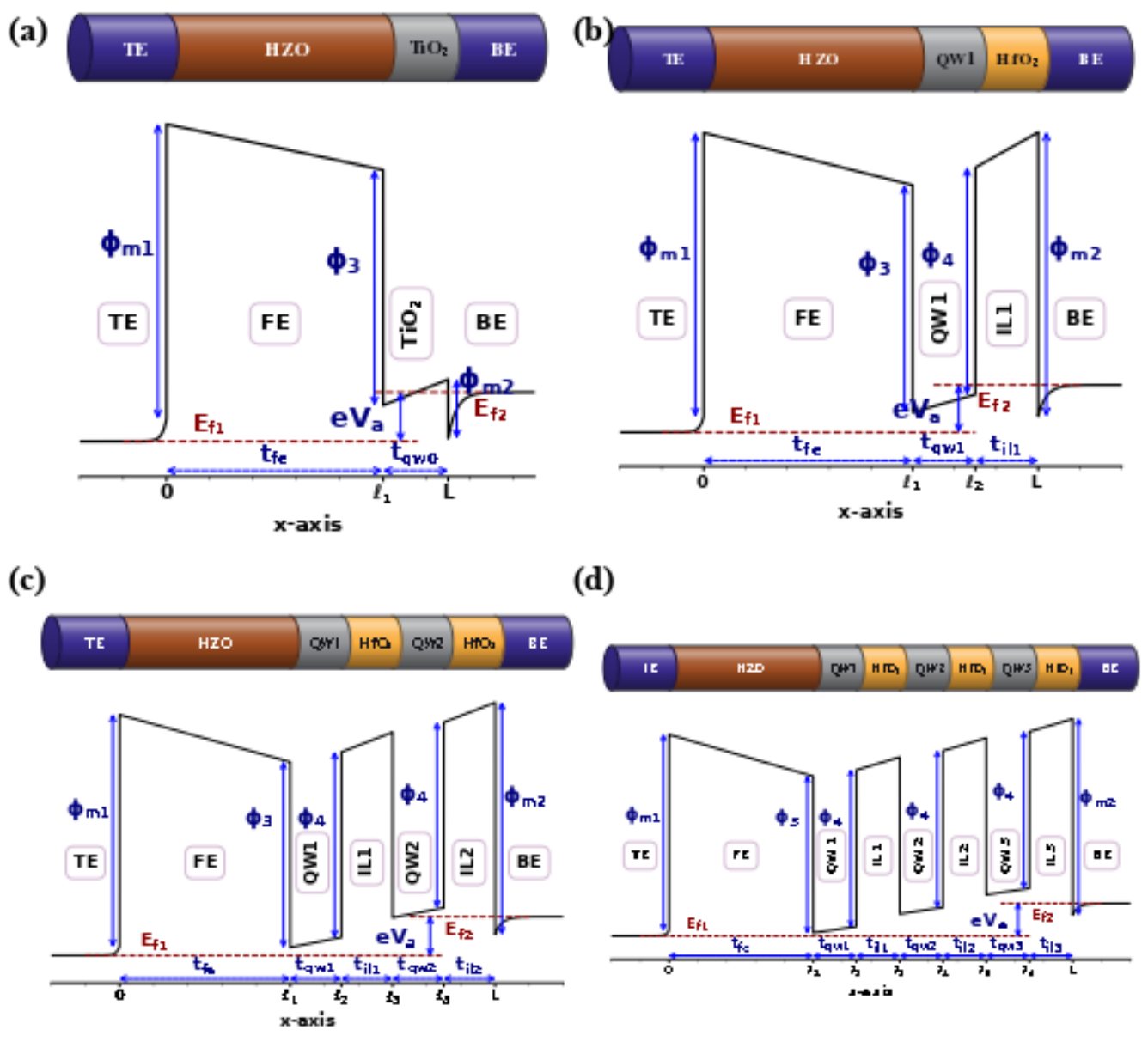

Fig. 1. Device schematics and corresponding energy-band profiles of TiN/HZO-based FTJs with different QW configurations: (a) 0-QW (FE/IL bilayer), (b) 1-QW, (c) 2-QW, and (d) 3-QW.

of finite electrode screening and elastic scattering on these characteristics also require further investigation.

In this work, multilayer $HfO_2$-based FTJs containing zero to three QWs are analyzed using the Miller phenomenological FE polarization model integrated with NEGF formalism. $TiO_2$ and IGZO-based QWs are first compared to examine how their dielectric and electronic properties affect polarization and the TER-RA characteristics. The $TiO_2$-based double-QW structure is then examined in detail. Its uniform configuration is used to investigate RT, polarization-dependent transport, and NDR. The effects of finite electrode screening on the TER-RA characteristics and elastic scattering on resonant transmission are also evaluated. Finally, the QW thicknesses are varied while keeping both IL thicknesses equal to examine how non-uniform QW thicknesses affect confined-state alignment, electron transport, and NDR. These results provide physical insight into how QW material and thickness, electrode screening, and elastic scattering affect polarization-dependent transport, and offer a theoretical basis for optimizing the TER–RA trade-off in multilayer FTJs.

## II. Device Modeling

The device configurations and their corresponding energy-band profiles are shown in Fig. 1. All structures consist of TiN top and bottom electrodes (TE and BE) and an HZO FE layer. An IL is retained in the 0-QW structure, whereas the other configurations contain 1, 2, or 3 QW–IL pairs. The general layer sequence is written as $\mathrm{M/FE/(QW/IL)}_n\mathrm{/M}$, where $n = 1, 2, 3$ represents the number of QW–IL pairs. The thicknesses and relative permittivities of the FE layer, $i$-th QW, and corresponding IL are denoted by $(t_{\mathrm{fe}}, \varepsilon_{\mathrm{fe}})$, $(t_{\mathrm{qw},i}, \varepsilon_{\mathrm{qw},i})$, and $(t_{\mathrm{il},i}, \varepsilon_{\mathrm{il},i})$, respectively. The electrostatic potential within the metal electrodes is described by the Thomas–Fermi screening model as follows: [2], [19], [20]:

$$V_{m1(m2)}(x) = \begin{cases} -\dfrac{\sigma_s \lambda_{m1}}{\varepsilon_0 \varepsilon_{m1}} e^{x/\lambda_{m1}}, & x \le 0, \\ \dfrac{\sigma_s \lambda_{m2}}{\varepsilon_0 \varepsilon_{m2}} e^{-(x-L)/\lambda_{m2}}, & x \ge L. \end{cases} \quad (2)$$

Here, $\sigma_s$ is the screening charge density, and $\lambda_{m1(m2)}$ and $\varepsilon_{m1(m2)}$ denote the screening lengths and effective relative permittivities of the TE and BE, respectively. The applied potential and built-in potential satisfy the following relation:

$$V_a + V_{\mathrm{bi}} = V_{m1}(0) + V_{m2}(L) + V_{\mathrm{fe}} + \sum_{i=1}^{n} (V_{\mathrm{qw},i} + V_{\mathrm{il},i}). \quad (3)$$

The built-in potential is given as:

$$V_{\mathrm{bi}} = \phi_{m2} - \phi_{m1} - i.(\phi_4 - \phi_3), \quad (4)$$

where $\phi_{m1(m2)}$ denote the electrode-interface barrier contributions, while $\phi_{3(4)}$ represent the internal-interface barrier contributions. Applying Gauss's law at the metal/FE interface and assuming continuity of the electric displacement field across the internal FE/QW and QW/IL interfaces give [21]

$$\sigma_s = P_{\mathrm{fe}} + \varepsilon_0 \varepsilon_{\mathrm{fe}} E_{\mathrm{fe}} = \varepsilon_0 \varepsilon_{\mathrm{qw},i} E_{\mathrm{qw},i} = \varepsilon_0 \varepsilon_{\mathrm{il},i} E_{\mathrm{il},i}, \quad (5)$$

The corresponding electric fields in the QW and IL layers are therefore

$$E_{\mathrm{qw},i} = \frac{\sigma_s}{\varepsilon_0 \varepsilon_{\mathrm{qw},i}}, \qquad E_{\mathrm{il},i} = \frac{\sigma_s}{\varepsilon_0 \varepsilon_{\mathrm{il},i}}. \quad (6)$$

The potential drop across the non-FE layers and electrode screening regions is

$$V_{\mathrm{S}} = \frac{\sigma_s}{\varepsilon_0} \left[ \sum_{i=1}^{n} \left( \frac{t_{\mathrm{qw},i}}{\varepsilon_{\mathrm{qw},i}} + \frac{t_{\mathrm{il},i}}{\varepsilon_{\mathrm{il},i}} \right) + \frac{\lambda_{m1}}{\varepsilon_{m1}} + \frac{\lambda_{m2}}{\varepsilon_{m2}} \right]. \quad (7)$$

For convenience, the combined electrical-thickness contribution of these layers and the electrode screening regions is denoted by $S$:

$$S = \sum_{i=1}^{n} \left( \frac{t_{\mathrm{qw},i}}{\varepsilon_{\mathrm{qw},i}} + \frac{t_{\mathrm{il},i}}{\varepsilon_{\mathrm{il},i}} \right) + \frac{\lambda_{m1}}{\varepsilon_{m1}} + \frac{\lambda_{m2}}{\varepsilon_{m2}}. \quad (8)$$

Thus, $V_S = \sigma_s S/\varepsilon_0$. Substituting the displacement-field relation and $V_{\mathrm{fe}} = E_{\mathrm{fe}} t_{\mathrm{fe}}$ into the voltage balance gives

$$V_a + V_{bi} = E_{fe} t_{fe} + \left( \frac{P_{\mathrm{fe}}}{\varepsilon_0} + \varepsilon_{\mathrm{fe}} E_{\mathrm{fe}} \right) S. \quad (9)$$

Solving for the voltage drop across the FE layer yields

$$V_{\mathrm{fe}} = \frac{t_{\mathrm{fe}} \left( V_a + V_{\mathrm{bi}} - \dfrac{P_{\mathrm{fe}} S}{\varepsilon_0} \right)}{t_{\mathrm{fe}} + \varepsilon_{\mathrm{fe}} S}. \quad (10)$$

The FE polarization is described using the phenomenological Miller model. Its voltage-dependent evolution is given by [22]

$$\frac{\partial P_{\mathrm{fe}}}{\partial V_{\mathrm{fe}}} = \delta_m \frac{\partial P_{\mathrm{sat}}}{\partial V_{\mathrm{fe}}}, \quad (11)$$

where $P_{fe}$ is the polarization along the applied-voltage trajectory, $P_{sat}$ is the saturated-branch polarization, and $\delta_m$ is the switching factor, defined as

$$\delta_m = 1 - \tanh \left[ \left( \frac{P_{\mathrm{fe}} - P_{\mathrm{sat}}}{\pm P_s - P_{\mathrm{fe}}} \right)^{1/2} \right], \quad (12)$$

TABLE I

FE AND MATERIAL PARAMETERS USED IN THE FTJ SIMULATIONS.

| HZO FE [15] | | TiN electrode | |
|---|---|---|---|
| Parameter | Value | Parameter | Value |
| $t_{fe}$ | 5 nm | $\Phi_m$ | 4.3 eV [26] |
| $E_c$ | 1.2 MV/cm | $m_e^*/m_o$ | 1.5 [27] |
| $P_r$ | 12 $\mu C/cm^2$ | $\varepsilon_r$ | 3 [28] |
| $P_s$ | 13 $\mu C/cm^2$ | $\lambda_m$ | 0.5 Å [14] |

| Material | Role | $m_e^*/m_o$ | $\chi$ (eV) | $\varepsilon_r$ |
|---|---|---|---|---|
| HZO | FE | 0.14 | 2.80 | 25 |
| $HfO_2$ | IL | 0.14 | 2.80 | 25 |
| $TiO_2$ | QW | 0.50 [29] | 4.00 [30] | 48 [31] |
| IGZO | QW | 0.34 [32] | 4.16 [33] | 10 [34] |

$t_{fe}$: FE thickness; $E_c$: coercive field; $m_e^*$: electron effective mass; $m_o$: free-electron mass; $\chi$: electron affinity; $\varepsilon_r$: relative permittivity; $\Phi_m$: metal work function; $\lambda_m$: metal screening length.

where $P_s$ is the saturation polarization. The saturated polarization is given by

$$P_{\text{sat}} = P_s \tanh\left[\frac{V_{\text{fe}} \mp V_c}{2V_c} \ln\left(\frac{P_s + P_r}{P_s - P_r}\right)\right], \quad (13)$$

where $P_r$ is the remanent polarization and $V_c$ is the coercive voltage. The upper and lower signs correspond to increasing and decreasing voltage sweeps, respectively. The polarization is obtained by numerically integrating Eq. (11) along the applied-voltage waveform. The current density is evaluated using the Landauer formalism [23]–[25]:

$$J = -\frac{2e}{Ah} \sum_{k_y,k_z} \int_{-\infty}^{+\infty} T_m(E) \left[f_{m1}(E) - f_{m2}(E)\right] dE, \quad (14)$$

where $e$ is the elementary charge, $A$ is the cross-sectional area, and $h$ is Planck's constant. Here, $T_m(E)$ denotes the transmission probability, while $f_{m1}(E)$ and $f_{m2}(E)$ are the Fermi–Dirac distributions of the TE and BE, respectively. Within the NEGF formalism, the total retarded and in-scattering self-energies are written as

$$\Sigma^r = \Sigma^r_{m1} + \Sigma^r_{m2} + \Sigma^r_o, \quad (15)$$

$$\Sigma^{\text{in}} = \Gamma_{m1} f_{m1} + \Gamma_{m2} f_{m2} + \Sigma^{\text{in}}_o, \quad (16)$$

where $\Sigma^r_{m1(m2)}$ are the retarded contact self-energies and $\Gamma_{m1(m2)}$ are the corresponding contact broadening matrices. Elastic scattering is incorporated through the retarded and in-scattering self-energies

$$\Sigma^r_o = D \circ G^r, \qquad \Sigma^{\text{in}}_o = D \circ G^n, \quad (17)$$

where $\circ$ represents element-wise multiplication, $D$ is the scattering-correlation matrix, $G^r$ and $G^n$ denote the retarded and electron correlation function, respectively. For local scattering, the matrix elements are given by

$$D_{pq} = \delta_{pq} D_0, \quad (18)$$

where $D_0$ controls the dephasing strength with $p$ and $q$ denoting local basis indices.

In the computational framework, the Miller model is integrated with the NEGF formalism. The FE polarization is evolved under the applied voltage using the fourth-order Runge–Kutta (RK4) method. At each RK4 stage, the polarization and applied voltage are used to calculate the screening charge density $\sigma_s$ and the voltage drop across the FE layer through the electrostatic relations in Eqs. (2)–(10). The resulting FE voltage is then used to evaluate the polarization derivative using the Miller model [Eqs. (11)–(13)]. The updated polarization and screening charge density determine the electric-field and electrostatic-potential profiles across the device layers. These profiles, along with the material-dependent band offsets, are used in the NEGF formalism to calculate the transmission and current density. The current density is subsequently obtained using the Landauer formalism [Eq. (14)]. Forward and reverse voltage sweeps provide the $P_{\text{fe}}$–$V_a$ hysteresis and $J$-$V_a$ characteristics for different QW–IL configurations. The effect of elastic scattering is also evaluated for the $TiO_2$-based 2-QW FTJ using the scattering formulation in Eqs. (15)–(18).

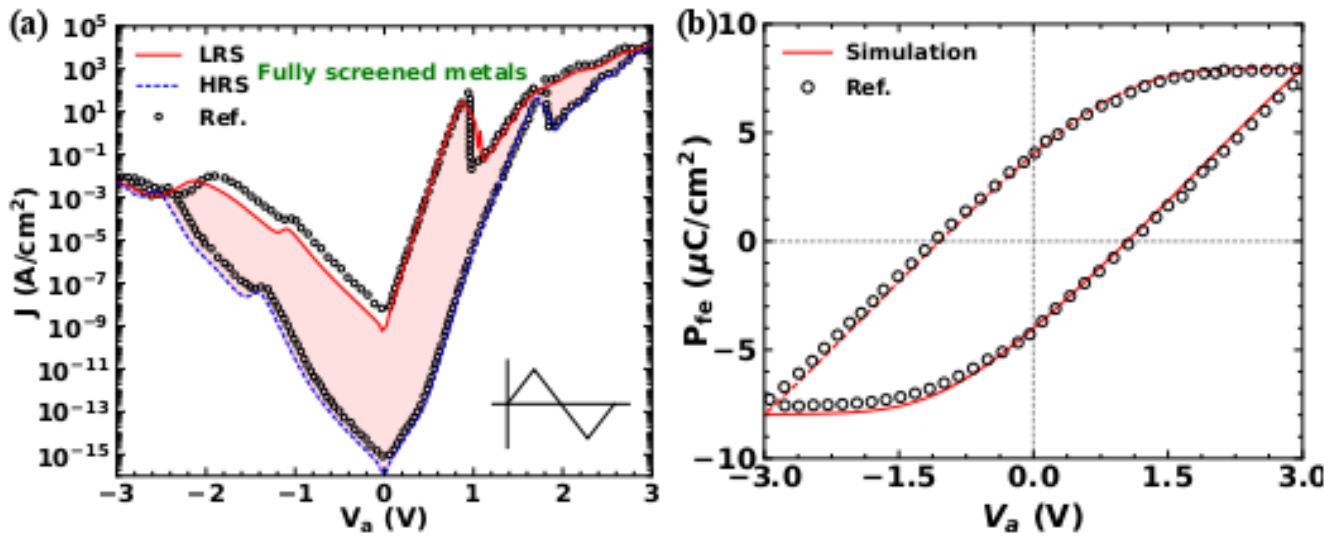

Fig. 2. Comparison of the simulated $J$-$V_a$ and $P_{fe}$–$V_a$ characteristics with the reference data reported in Ref. [15] for a double-QW FTJ with a $TiN/HZO/(Ta_2O_5/HfO_2)_2/TiN$ stack.

The simulation parameters are listed in Table I. The framework is validated by comparing the calculated $J$-$V_a$ and $P_{\text{fe}}$–$V_a$ characteristics with the reference results reported in Ref. [15] for a double-QW FTJ with a $TiN/HZO/(Ta_2O_5/HfO_2)_2/TiN$ stack, as shown in Fig. 2. The simulated polarization loop reproduces the overall hysteretic behavior of the reference data, while the corresponding current characteristics show good agreement, supporting the consistency of the polarization, electrostatic, and quantum-transport calculations.

## III. Results and Discussion

### A. Effect of QW Number and Material

To investigate the influence of QW number and material on electron transport, multilayer FTJs incorporating $TiO_2$ and IGZO-based QWs are compared. Finite screening effects at both the TE and BE are considered, with a screening length of $\lambda_m$=0.5 Å. The corresponding $J$-$V_a$ characteristics under a triangular voltage waveform are shown in Fig. 3(a) and Fig. 3(b), respectively. Both materials exhibit changes in current density with QW number, reflecting modifications in the RT pathways introduced by the QW–IL layers. To examine the changes in transport with QW number, the corresponding zero-bias transmission spectra are shown in Fig. 3(c) and Fig. 3(d). These spectra exhibit resonant peaks associated with QW-confined states. The energy separation, $\Delta E$, quantifies the difference between the energy positions of the first resonant peaks in the LRS and HRS. For one, two, and three QWs, $\Delta E$ decreases from 0.59 to 0.50 and 0.42 eV for $TiO_2$, and from 0.48 to 0.33 and 0.22 eV for IGZO. $TiO_2$-based FTJs exhibit a larger peak separation than IGZO-based structures for each QW configuration, indicating a greater spectral distinction between the two resistance states. The RA products are shown in Fig. 3(e) and Fig. 3(f) for $TiO_2$ and IGZO-based FTJs, respectively. At a read bias of 0.1 V, the $TiO_2$-based 2-QW FTJ exhibits an RA product of $5 \times 10^4\ \Omega \cdot \mathrm{cm}^2$, compared with $2 \times 10^5\ \Omega \cdot \mathrm{cm}^2$ for the IGZO-based structure. The corresponding TER characteristics are presented in Fig. 3(g) and Fig. 3(h), respectively. At the same read bias, the TER is $2.71 \times 10^{10}\%$ for $TiO_2$ and $5.95 \times 10^7\%$ for IGZO. Thus, the $TiO_2$-based double-QW structure provides both a lower RA product and a higher TER than its IGZO counterpart. This difference is related to the dielectric and electronic properties of the QW materials. Although $TiO_2$ has a slightly lower electron affinity (4.00 eV) than IGZO (4.16 eV), its dielectric permittivity is considerably higher. The higher permittivity results in a lower electric field within the QW and modifies the voltage distribution and electrostatic potential profile across the multilayer stack. Meanwhile, differences in electron affinity and effective mass influence the band alignment and quantum-confinement energies, respectively. These changes affect the confined-state energies and their alignment in the LRS and HRS, thereby modifying resonant transmission and current modulation.

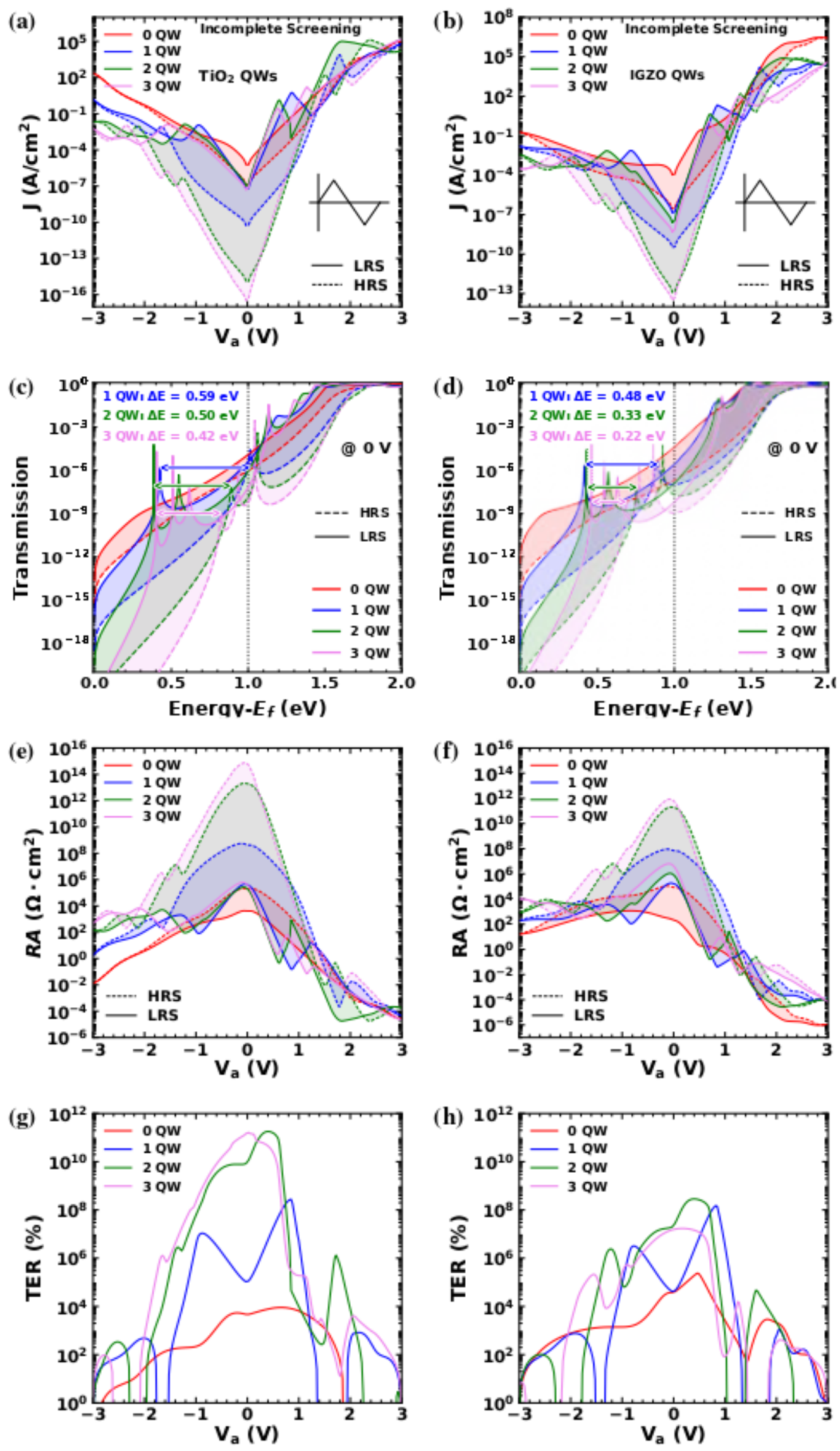

Fig. 3. Effect of QW number on the transport characteristics of multilayer FTJs incorporating $TiO_2$-based (left panels) and IGZO-based (right panels) QWs. (a), (b) $J$-$V_a$ characteristics; (c), (d) transmission spectra; (e), (f) RA products; and (g), (h) TER.

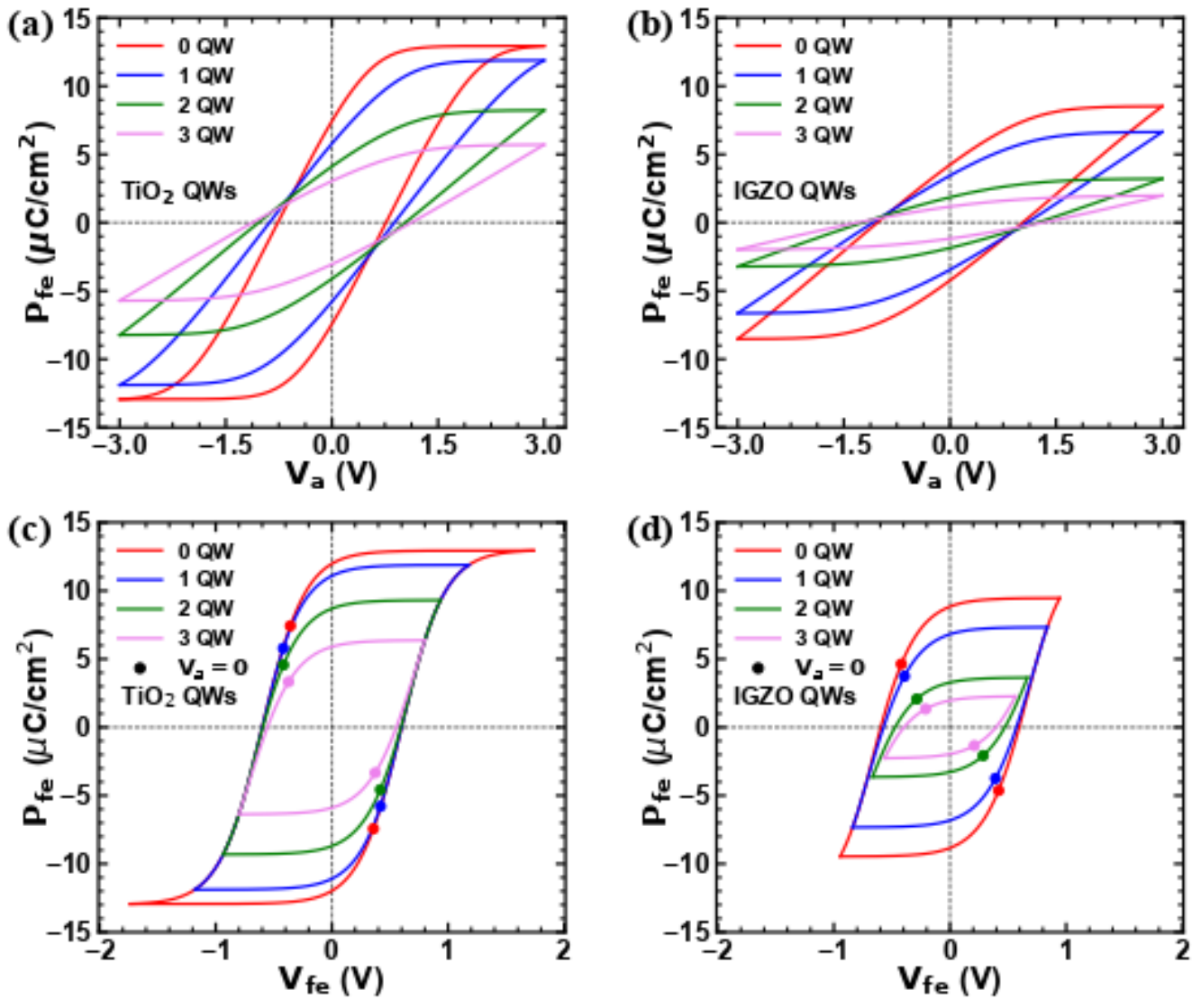

Fig. 4. Polarization characteristics of multilayer FTJs incorporating $TiO_2$-based (left panels) and IGZO-based (right panels) QWs with different QW numbers. (a), (b) $P_{fe}$–$V_a$ hysteresis loops; (c), (d) $P_{fe}$–$V_{fe}$ characteristics.

The polarization hysteresis loops in Fig. 4(a) and Fig. 4(b) compare $TiO_2$ and IGZO-based FTJs, respectively, with 0–3 QWs. $TiO_2$-based structures retain a larger remanent polarization as compared to IGZO-based structures, while increasing the QW number reduces the polarization in both materials. This trend originates from the increased effective electrical thickness of the non-FE layers, $S$, which modifies the voltage distribution [Eqs. (8) and (9)] and reduces the voltage available across the FE layer [Eq. (10)]. The corresponding polarization versus FE-layer voltage is presented in Fig. 4(c) and Fig. 4(d). The non-zero polarization at $V_{\mathrm{fe}}$=0 reflects the persistence of polarization under the internal depolarizing field. At zero voltage, the magnitudes of the depolarizing field and voltage are expressed as

$$|E_{\mathrm{dep},0}| = \frac{|P_{\mathrm{fe}}|S}{\varepsilon_0(t_{\mathrm{fe}} + \varepsilon_{\mathrm{fe}}S)}, \quad (19)$$

$$|V_{\mathrm{dep},0}| = \frac{|P_{\mathrm{fe}}|t_{\mathrm{fe}}}{\varepsilon_0\varepsilon_{\mathrm{fe}}} \frac{S}{S + t_{\mathrm{fe}}/\varepsilon_{\mathrm{fe}}}. \quad (20)$$

The depolarizing field depends on both the electrical thickness $S$ and the FE polarization $P_{\mathrm{fe}}$. Increasing the QW number increases the electrical thickness, which tends to enhance the depolarizing field. At the same time, the reduction in polarization tends to lower its magnitude. Thus, the net depolarizing field is determined by the combined variation of $S$ and $P_{fe}$. The material dependence arises from the dielectric contribution of the QWs. For equal physical thickness, $t_{qw}/\varepsilon_{TiO_2} < t_{qw}/\varepsilon_{IGZO}$, since $\varepsilon_{TiO_2}$=48 and $\varepsilon_{IGZO}$=10. For the same polarization and remaining device parameters, the smaller electrical-thickness contribution of $TiO_2$ results in a lower depolarizing field and voltage, thereby favoring stronger polarization retention.

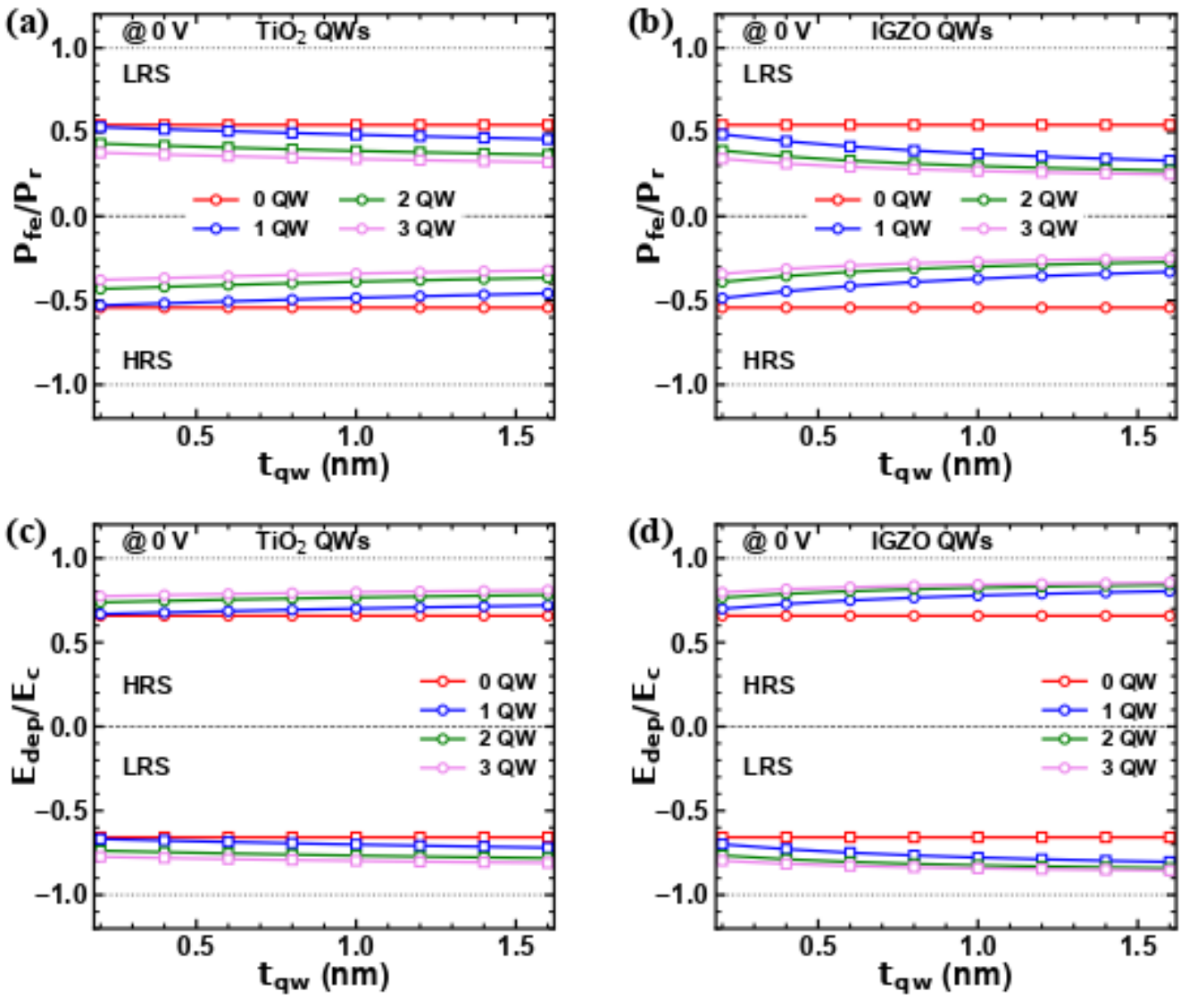

Fig. 5. QW-thickness dependence of normalized polarization and depolarizing field at zero applied voltage for FTJs incorporating $TiO_2$-based (left panels) and IGZO-based (right panels) QWs. (a), (b) Normalized polarization, $\boldsymbol{P_{\mathrm{fe}}/P_r}$; (c), (d) normalized depolarizing field, $\boldsymbol{E_{\mathrm{dep}}/E_c}$. Results are shown for 0-3 QWs, with LRS and HRS distinguished by different line styles.

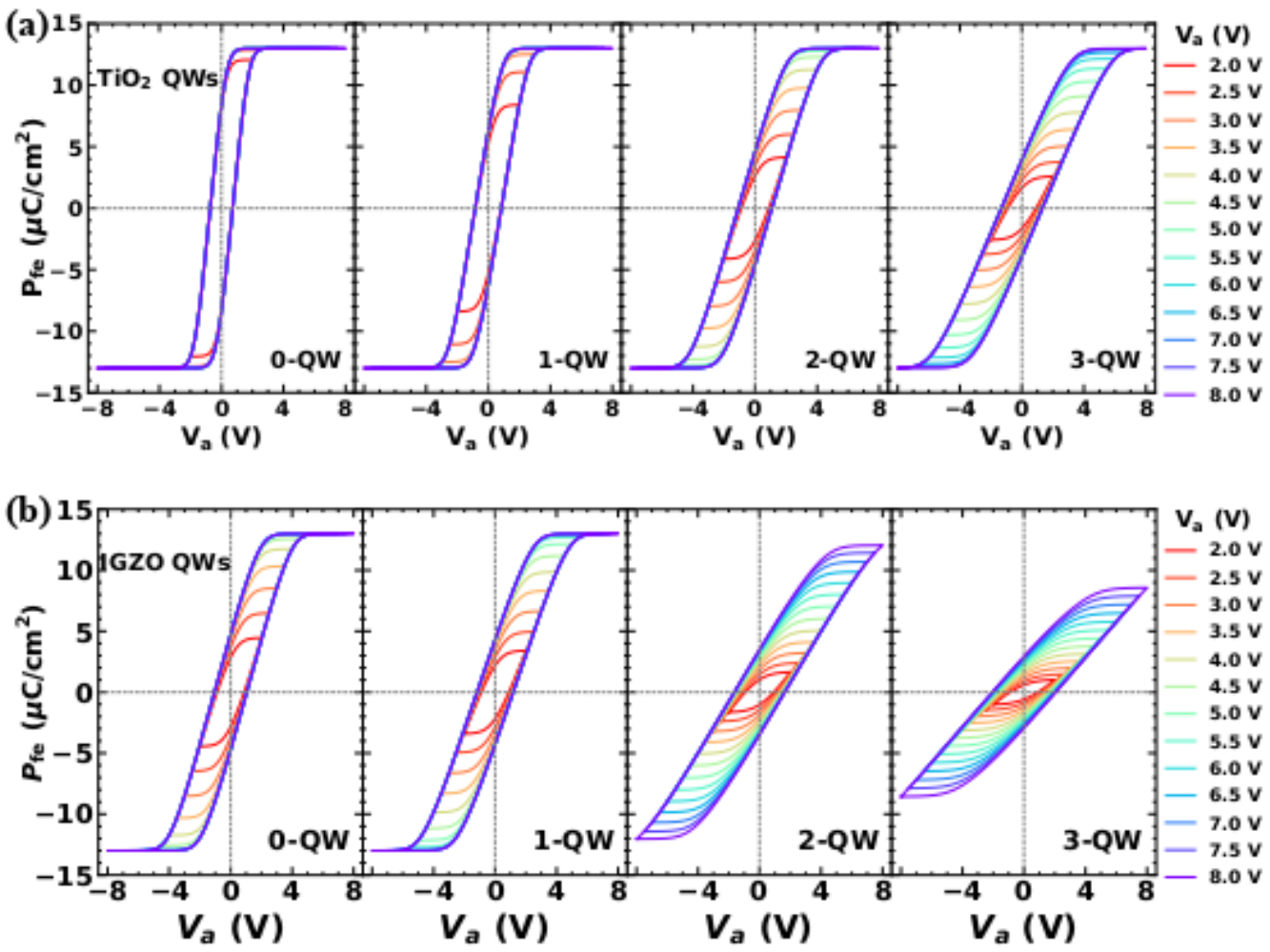

Fig. 6. Applied-voltage-dependent polarization hysteresis of multilayer FTJs incorporating (a) $TiO_2$-based (top row) and (b) IGZO-based (bottom row) QWs. The columns from left to right correspond to 0-3 QWs.

Fig. 5 shows the QW-number and thickness dependence of normalized polarization ($P_{\mathrm{fe}}/P_r$) and depolarizing field ($E_{\mathrm{dep}}/E_c$) at zero applied voltage for $TiO_2$ and IGZO-based FTJs. For $TiO_2$-based FTJs, Fig. 5(a) shows that the normalized polarization decreases with increasing QW number, while a slight reduction is observed with increasing uniform QW thickness ($t_{qw1}$=$t_{qw2}$=$t_{qw}$). In comparison, IGZO-based FTJs exhibit a slightly greater reduction in normalized polarization with increasing QW number and thickness, as shown in Fig. 5(b). The corresponding normalized depolarizing-field magnitude in Fig. 5(c) increases with QW number but remains almost unchanged with increasing QW thickness. For IGZO-based FTJs, the normalized depolarizing-field magnitude also increases with QW number and exhibits slightly higher values than those of $TiO_2$-based structures, as shown in Fig. 5(d). This material-dependent behavior is associated with the dielectric permittivities of the QW materials. Since $\varepsilon_{IGZO}$=10 is lower than $\varepsilon_{TiO_2}$=48, IGZO contributes more electrical thickness for the same physical QW thickness, resulting in a higher depolarizing field. In contrast, the smaller electrical-thickness contribution of $TiO_2$ favors better polarization retention.

The applied-voltage dependent polarization hysteresis for different QW configurations is compared in Fig. 6(a) and Fig. 6(b) for $TiO_2$ and IGZO-based FTJs, respectively. For both materials, increasing the QW number extends the polarization-switching region over a wider voltage range, indicating that higher applied voltages are required to reverse the polarization. For a given QW configuration, increasing the maximum applied-voltage amplitude drives the polarization closer to saturation. However, incomplete polarization switching is more evident in IGZO-based structures with higher QW numbers, where saturation is not reached even at the maximum applied-voltage amplitude. In comparison, $TiO_2$-based structures approach polarization saturation at comparatively lower voltage amplitudes. These differences are associated with the larger effective electrical-thickness contribution of IGZO, which reduces the voltage available across the FE layer and consequently limits polarization switching. Considering the combined polarization and transport characteristics discussed above, the $TiO_2$-based 2-QW FTJ is selected for further investigation of its transport mechanisms and structural optimization.

### *B. $TiO_2$-based 2-QW FTJ*

*1) Uniform QW and IL Layers:* The transport characteristics of the $TiO_2$-based 2-QW FTJ with 1-nm-thick QWs and ILs are shown in Fig. 7. The $J$-$V_a$ characteristics in Fig. 7(a) show distinct current densities in the LRS and HRS. The LRS current also exhibits NDR, which is associated with RT through QW-confined states. As shown by the bias-dependent transmission spectra in Fig. 7(b), resonant transmission occurs at $V_{reso}$ = 0.6 V when the confined states align with the electrode transport-energy window. This alignment enhances electron transmission and increases the current. With further increase in bias, the confined states shift away from the transport-energy window, reducing the resonant contribution to the current and resulting in NDR. The zero-bias transmission spectra in Fig. 7(c) further reveal different transmission probabilities for the LRS and HRS, resulting from polarization-dependent changes in the tunneling barrier. The corresponding RA and TER characteristics are shown in Fig. 7(d). At a read bias of $V_r$=0.1 V, the uniform double-QW structure exhibits a TER of $2.71 \times 10^{10}\%$ and an RA product of $5 \times 10^4\ \Omega \cdot \mathrm{cm}^2$.

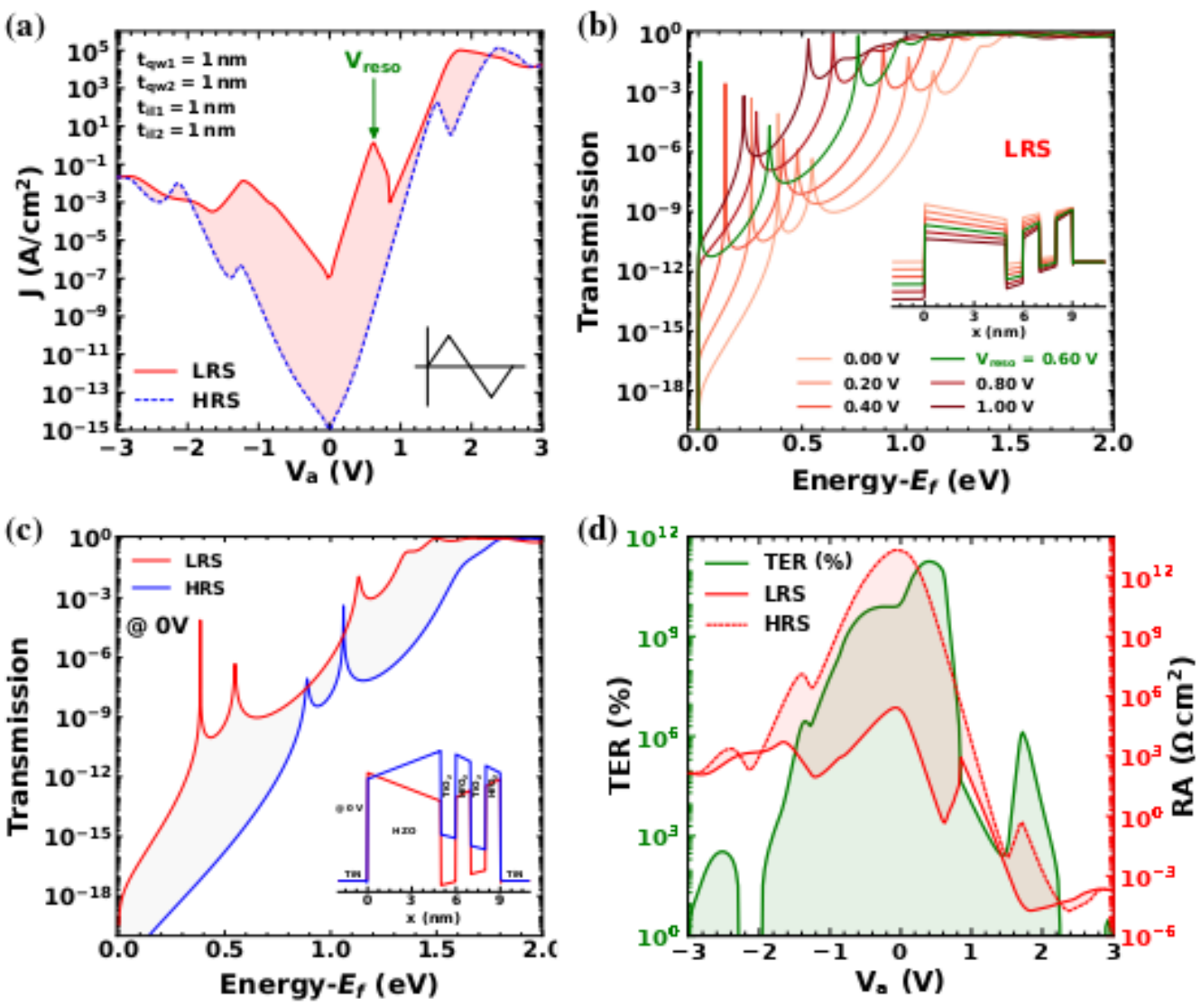

Fig. 7. Transport characteristics of the $TiO_2$-based 2-QW FTJ with 1 nm thick QWs and ILs: (a) $J$-$V_a$ characteristics; (b) bias-dependent transmission spectra showing resonant transmission at $V_{reso}$=0.6 V; (c) zero-bias transmission spectra; and (d) bias-dependent TER and RA product.

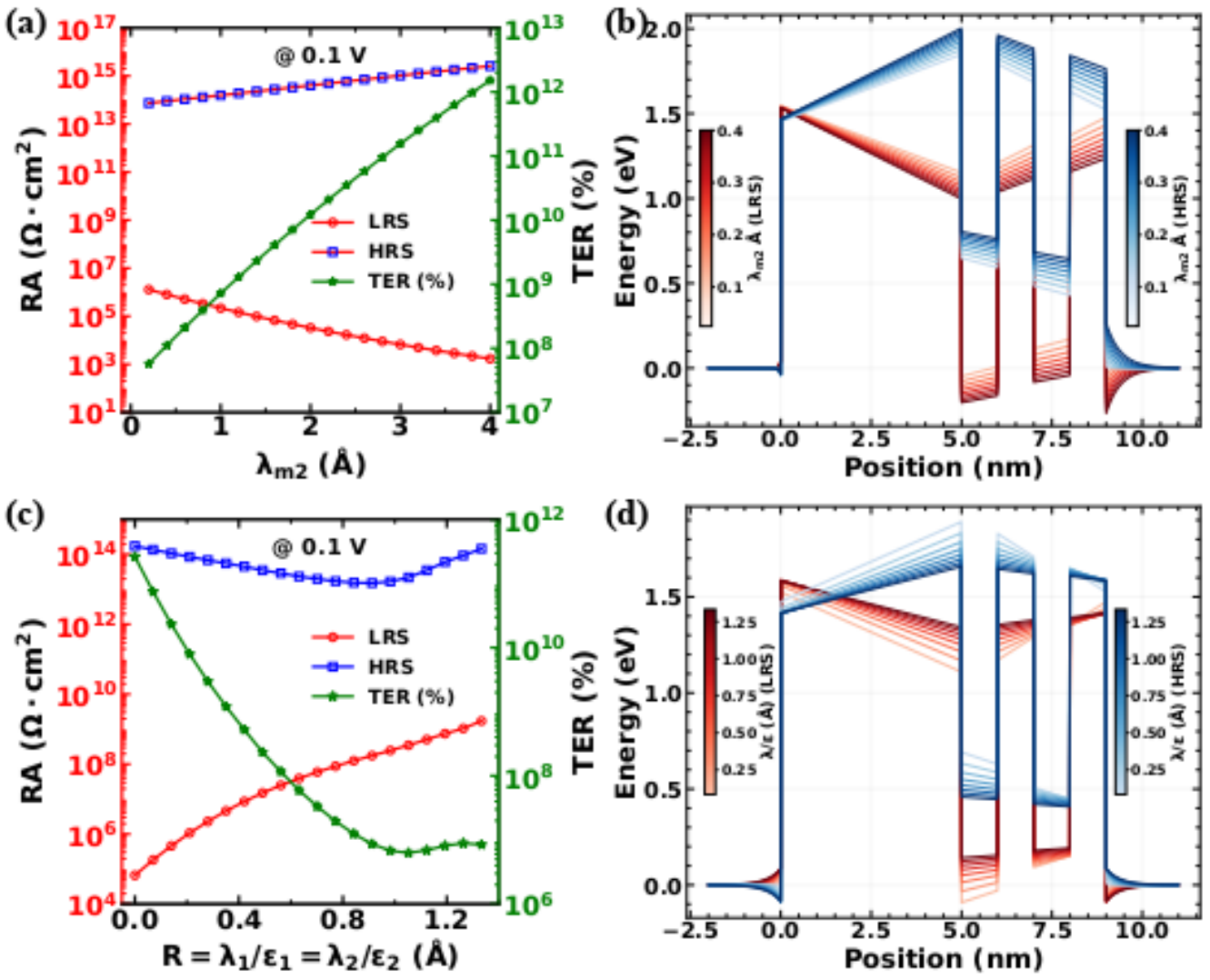

Fig. 8. Effect of electrode screening on the transport characteristics and energy-band profiles of the $TiO_2$-based 2-QW FTJ. (a), (b) RA products and TER as functions of the screening length $\lambda_{m2}$ and the corresponding energy-band profiles, respectively; (c), (d) RA products and TER as functions of the screening parameter $R$=$\lambda_m/\varepsilon_m$ and the corresponding energy-band profiles, respectively. All remaining simulation parameters are kept unchanged.

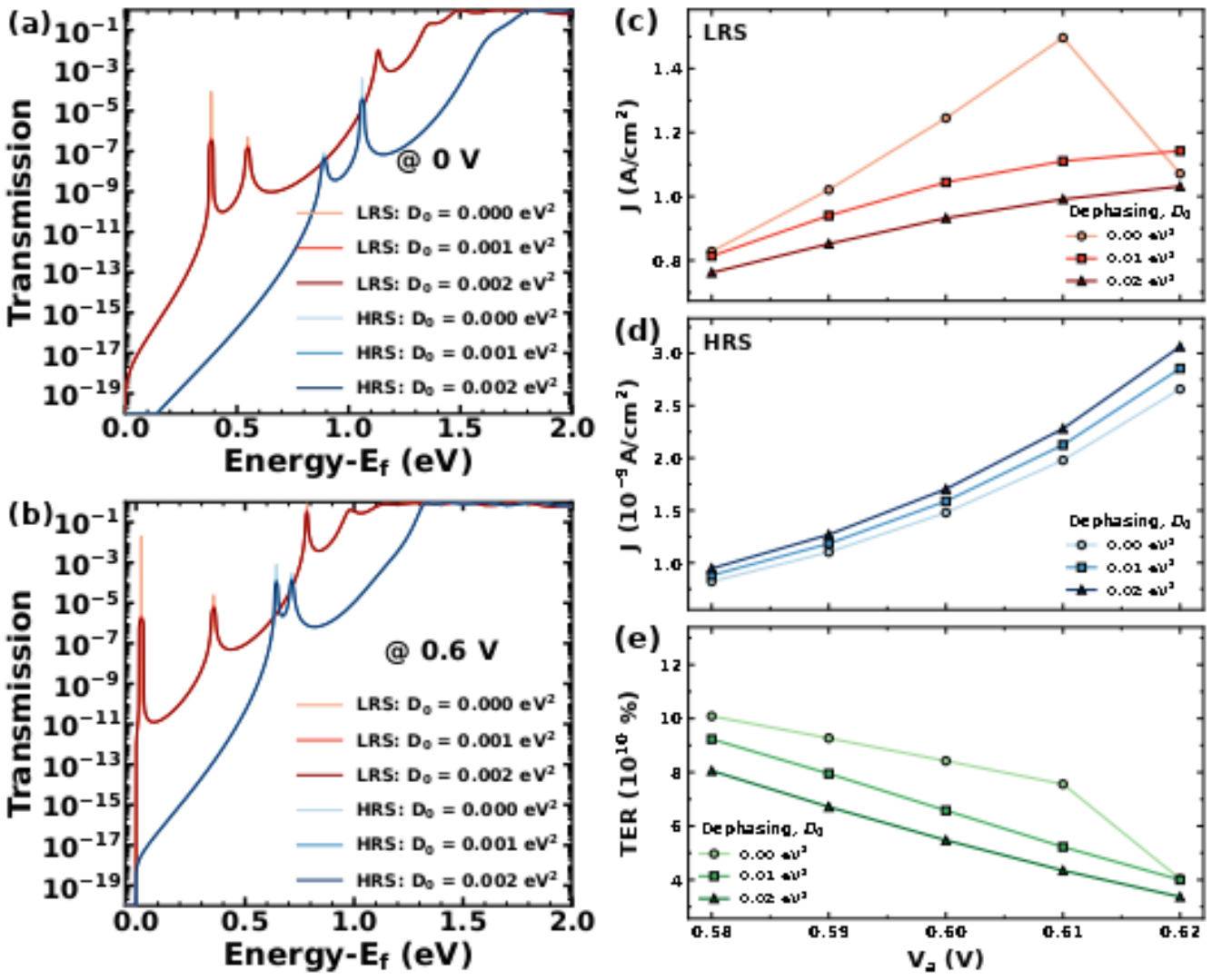

Fig. 9. Effect of elastic dephasing strength $D_0$ on the transport characteristics of the $TiO_2$-based 2-QW FTJ. Transmission spectra of the LRS and HRS at $V_a$ (a) 0 V and (b) 0.6 V. Bias-dependent RA products for (c) the LRS and (d) the HRS, and (e) the corresponding TER for different $D_0$ values.

The influence of electrode screening on the transport and electrostatic properties of the $TiO_2$-based 2-QW FTJ is examined in Fig. 8. As shown in Fig. 8(a), increasing the screening length $\lambda_{m2}$ at $V_r$=0.1 V reduces the LRS-RA product while increasing the HRS-RA, resulting in a higher TER. The corresponding energy-band profiles in Fig. 8(b) illustrate the changes in the potential distribution and QW energy levels with screening length. The effect of the normalized screening parameter, $R = \lambda_m/\varepsilon_m$, is shown in Fig. 8(c). With increasing $R$, the LRS-RA increases, whereas the HRS-RA initially decreases and then increases slightly at larger $R$. Accordingly, the TER decreases over most of the investigated range. The corresponding band profiles in Fig. 8(d) show the associated changes in the potential profile and QW confinement. These results demonstrate that electrode screening modifies the electrostatic potential profile and, consequently, the tunneling characteristics and current modulation between the resistance states.

The effect of elastic scattering on resonant transport in the $TiO_2$-based 2-QW FTJ is examined in Fig. 9. The transmission spectra at $V_a = 0$ V and 0.6 V in Fig. 9(a) and Fig. 9(b), respectively, show that increasing the scattering strength $D_0$ weakens and broadens the resonant peaks associated with QW states due to reduced phase coherence. The resulting transmission within the bias window determines the current densities $J_{LRS}$ and $J_{HRS}$, shown in Fig. 9(c) and Fig. 9(d), respectively. $J_{LRS}$ exhibits a stronger dependence on $D_0$, particularly at higher bias, whereas the $J_{HRS}$ curves remain relatively close. Consequently, the TER varies with both bias and scattering strength, as shown in Fig. 9(e). The TER decreases with increasing bias over the investigated range, indicating a reduction in the difference between the two resistance states. Overall, elastic scattering affects LRS and HRS transport differently, thereby influencing the TER characteristics.

*2) Non-Uniform QW Layers:* The effect of non-uniform QW thickness on the transport characteristics of the $TiO_2$-based 2-QW FTJ is investigated by varying $t_{\mathrm{qw1}}$ for different values of $t_{\mathrm{qw2}}$. The LRS and HRS RA products at $V_r$=0.1 V for the investigated thickness combinations are presented in Fig. 10(a), while the corresponding TER is shown in Fig. 10(b). The TER reaches its maximum at $t_{\mathrm{qw1}}$=1.38 nm. This dependence arises from changes in quantum confinement and the relative alignment of QW-confined states, which affect the LRS and HRS transmission probabilities differently. Considering the TER enhancement together with the corresponding RA values, the configuration with $t_{\mathrm{qw1}}$=1.38 nm, $t_{\mathrm{qw2}}$=1 nm, and $t_{\mathrm{il1}}$=$t_{\mathrm{il2}}$=1 nm is selected as the optimum TER-RA configuration. The corresponding $J$-$V_a$ characteristics in Fig. 10(c) show distinct current densities for the LRS and HRS, along with an NDR feature at $V_{reso}$=0.36 V. Similar to the uniform structure, this behavior is associated with RT through QW-confined states. The bias-dependent transmission spectra in Fig. 10(d) show changes in resonant transmission as the confined-state energies shift relative to the electrode transport-energy window. The zero-bias transmission spectra in Fig. 10(e) also reveal different transmission probabilities for the LRS and HRS due to polarization-dependent barrier modulation. These differences are reflected in the voltage-dependent TER and RA characteristics in Fig. 10(f). The selected non-uniform double-QW structure achieves a TER of $2.95 \times 10^{11}$% and an RA product of $1.25 \times 10^3\ \Omega \cdot \mathrm{cm}^2$ at $V_r$=0.1 V. Thus, it provides a more favorable TER-RA trade-off, with approximately 10.9 times higher TER and 40 times lower RA than the uniform double-QW structure. This improvement results from the unequal QW thicknesses, which modify the confined-state alignment and polarization-dependent transmission in the two resistance states.

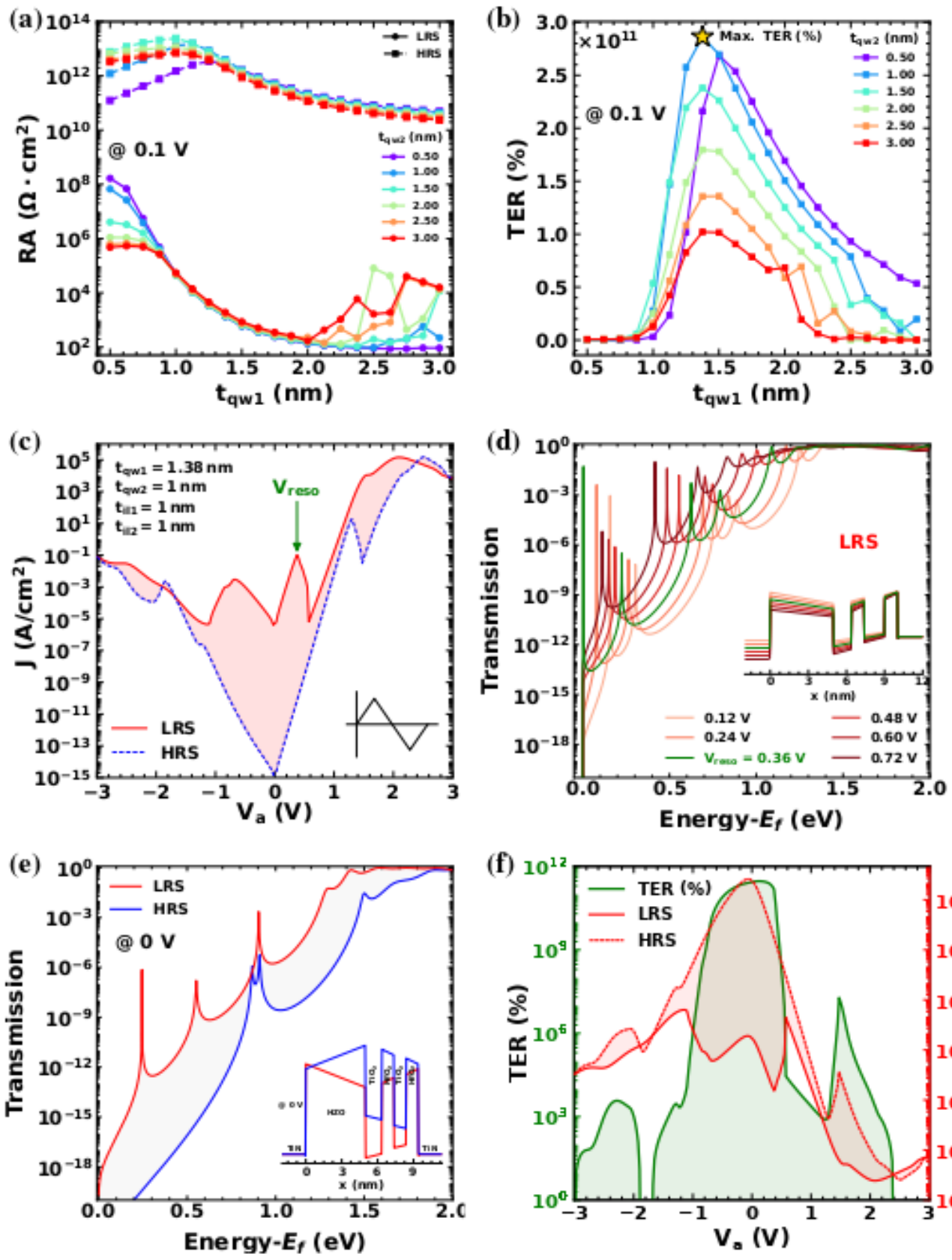

Fig. 10. Effect of non-uniform QW thickness on the transport characteristics of the $TiO_2$-based 2-QW FTJ. (a) LRS and HRS RA products and (b) corresponding TER as functions of $t_{\mathrm{qw1}}$ for different values of $t_{\mathrm{qw2}}$, evaluated at $V_r$=0.1 V. (c) $J$-$V_a$ characteristics of the selected geometry ($t_{qw1}$=1.38 nm, $t_{qw2}$=1 nm, and $t_{il1}$=$t_{il2}$=1 nm); (d) bias-dependent transmission spectra showing resonant transmission near $V_{reso} = 0.36$ V; (e) zero-bias transmission spectra for the LRS and HRS; and (f) bias-dependent TER and RA products.

## IV. Conclusion

In this work, the Miller phenomenological polarization model is coupled with the NEGF formalism to investigate the electrostatics and quantum transport characteristics of multilayer $HfO_2$-based FTJs. A comparison of $TiO_2$ and IGZO-based QWs shows that their dielectric and electronic properties strongly influence polarization retention and TER-RA characteristics. $TiO_2$-based structures exhibit higher polarization retention and TER than IGZO-based structures, with the double-QW configuration offering the most favorable TER–RA trade-off. The transport analysis further shows that RT and NDR originate from transport through QW-confined states. In addition, finite electrode screening modifies the electrostatic potential profile, while elastic dephasing affects resonant transmission. The investigation of non-uniform QW thicknesses further demonstrates that variations in individual QW thicknesses modify the confined-state energies and their relative alignment, thereby influencing the resistance-state characteristics and TER-RA trade-off. These findings show how QW material selection and thickness configuration influence polarization-dependent transport and the resulting TER–RA characteristics of multilayer FTJs.